# Large-area implementation and critical evaluation of the material and fabrication aspects of a thin-film thermoelectric generator based on aluminum-doped zinc oxide

Kirsi Tappura[a*], Taneli Juntunen[b], Kaarle Jaakkola[a], Mikko Ruoho[b], Ilkka Tittonen[b], Riina Ritasalo[c] and Marko Pudas[c]

[a]VTT Technical Research Centre of Finland Ltd., FI-02044 Espoo, Finland
[b]Department of Electronics and Nanoengineering, Aalto University, P.O. Box 13500, FI-00076 Aalto, Finland
[c]Picosun Oy, Tietotie 3, 02150 Espoo, Finland

*to whom correspondence should be addressed

E-mail address of the corresponding author:
kirsi.tappura@vtt.fi

## Abstract

A large-area thermoelectric generator (TEG) utilizing a folded thin-film concept is implemented and the performance evaluated for near room temperature applications having modest temperature gradients (< 50 K). The TEGs with the area of ~0.33 m$^2$ are shown capable of powering a wireless sensor node of multiple sensors suitable e.g. for monitoring environmental variables in buildings. The TEGs are based on a transparent, non-toxic and abundant thermoelectric material, i.e. aluminium-doped zinc oxide (AZO), deposited on flexible substrates. After folding, both the electrical current and heat flux are in the plane of the thermoelectric thin-film. Heat leakage in the folded TEG is shown to be minimal (close to that of air), enabling sufficient temperature gradients without efficient heat sinks, contrary to the conventional TEGs having the thermal flux and electrical current perpendicular to the plane of the thermoelectric films. The long-term stability studies reveal that there are no significant changes in the electrical or thermoelectric properties of AZO over several months, while the contact resistance between AZO and silver ink is an issue exhibiting a continuous increase over time. The performance of the TEGs and technological implications in relation to a state-of-the-art thermoelectric material are further assessed via a computational study.

Keywords: large-area thermoelectric generator; thin-film TEG; aluminum-doped zinc oxide; finite element method; atomic layer deposition

## 1. Introduction

Thermoelectric generators (TEGs) provide a means to produce useful electrical energy from waste heat that would otherwise be lost. They can generate energy in extreme environments and in remote areas where no power grids are available and where the battery replacement would be difficult and expensive. In addition to the extreme or remote environments, TEGs are attractive for autonomous sensor systems where the constant battery replacement would not only cause breaks in the operation but also destroy the idea of a truly autonomous system and increase the maintenance costs.

Although the properties of thermoelectric (TE) materials and the efficiency of TEGs have significantly improved as a result of recent research efforts [1, 2, 3], the efficiency of the thermal to electrical energy conversion is still much lower than that of the Carnot cycle [4]. A major challenge in the application of TEGs is, thus, their low energy conversion efficiency, which has led to a high cost per converted power [5]. This also relates to the fact that the commercial TEGs are mainly based on materials of high price and low abundance, such as tellurium and bismuth [6]. An additional issue is their toxicity.

An alternative approach is to concentrate on non-toxic and abundant materials providing reasonable performance and a promise for a reasonable cost even if their performance may not reach the level of the high-performance TE materials. As the awareness of environmental issues has grown over the past several years, increasing attention has been payed to boosting the thermoelectric properties of metal oxides that are seen as a group of promising alternative TE materials. Although tin-doped indium oxide (ITO), the commonly used transparent conductive oxide, has high electrical conductivity and high transparency, it cannot be considered as a sustainable choice. Instead, aluminium-doped zinc oxide (Al:ZnO or AZO) has been proposed as an environmentally friendly option that is not only non-toxic and more abundant but it also has better thermoelectric properties, high transparency, reasonable cost and can be used in low temperature applications [7, 8, 9, 10, 11].

The high cost per converted power and the material issues of the conventional TEGs create a major obstacle for the wider use of thermoelectric generators in different applications. This applies especially to large-area systems for which the TEG concept should be scalable, economical and, in the long term, based on sustainable materials. Therefore, even the technological demonstrations of large-area thermoelectric generators remain scarce for the time being. To the best of our knowledge, there are only two experimental demonstrations of truly large-area TEG systems: one based on electrospun silicon nanotubes as the TE nanomaterials [12] and another based on electrochemical deposition of Ni/Cu couples in perforated printed circuit board substrates [13].

Most of the thermal gradients around us, e.g. in buildings, are small and the usage of efficient heatsinks is expensive, inconvenient or impossible, which effectively prevents the employment of the conventional TEGs in such low-energy-density applications. A novel folded thin-film TEG concept suitable for large-area, heatsink-limited conditions has previously been introduced and the performance investigated computationally [14] as well as its advantages compared with the other TEG concepts and related studies by other groups. The preliminary experimental results of the folded TEG have also been reported in the context of performing as a power source for a window-integrated wireless sensor node with the TEGs measured in planar [15] and single module formats [16]. The principle of the TEG concept [14] and schematics of the folded TEGs before and after folding, as well as a picture of a fabricated TEG module of eleven TEG elements are shown in Fig. 1.

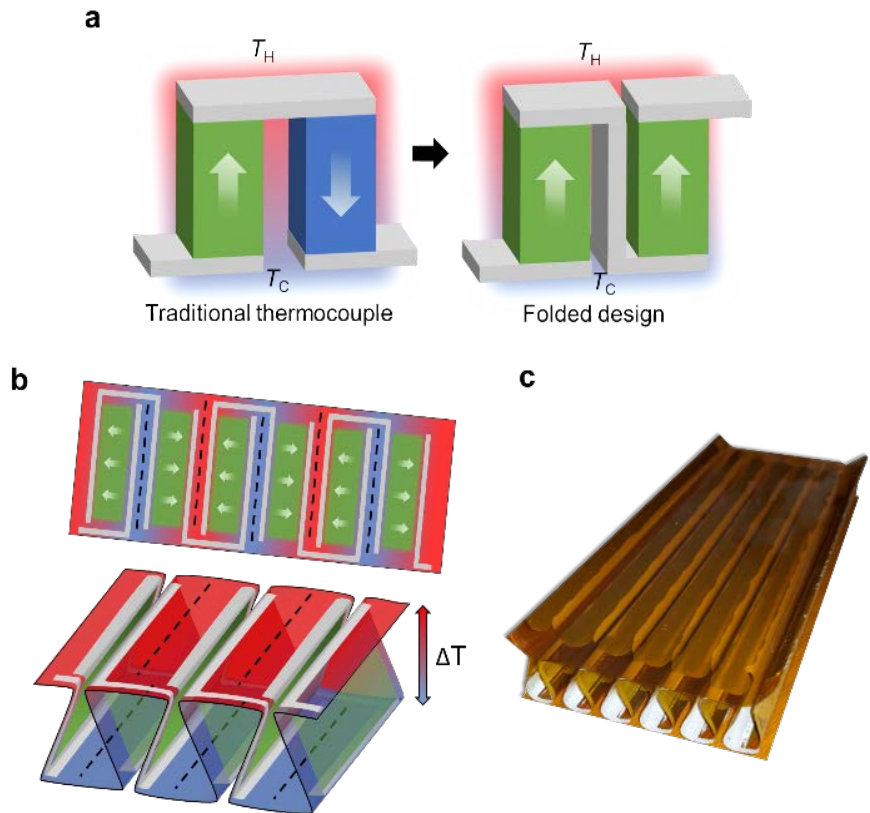

**Figure 1.** The folded TEG design: (a) A traditional TE couple compared with a TEG based on a single-type thermoelectric material used in the folded design. (b) A schematic illustration of the three-dimensional structure of the extended (above) and folded (below) TEG module. The color-coding indicates the temperature gradient (ΔT) over the module (blue = "cold", red = "hot") and the effective TE legs (green) formed by the current flow. The white arrows on green indicate the direction of the electrical current in AZO, i.e. in an n-type TE material. (c) A picture of the folded TEG module consisting of eleven TEG elements connected electrically in series and thermally in parallel.

In this paper, the folded thin-film TEG prepared on flexible low-cost substrates for heatsink-limited applications is fabricated in a large-area format (~0.33 m$^2$) exploiting AZO, the abundant environmentally friendly TE material grown by atomic layer deposition (ALD). It is shown, among other things, that a significant benefit of the proposed TEG design is that the heat leakage through the module itself can be minimized and the temperature gradient available for power conversion maximized. In addition to the characterization of the as-prepared AZO and electrical contacts, device simulations combined with the device characterization are applied to extract the material properties of the fabricated TEGs in their different fabrication stages enabling the identification of the most critical fabrication steps. All the material, design and fabrication aspects relevant to the large-area thin-film TEGs are critically evaluated in order to pave the way for future improvements.

## 2. Materials and methods

### 2.1. Material deposition and characterization

Al-doped ZnO (AZO) was grown on 25 μm thick polyimide (DuPont Kapton® NH) substrates by atomic layer deposition (ALD) using Picosun$^{TM}$ P-300B ALD reactor. The ALD process was selected to achieve conformal coatings at low process temperatures as required by the flexible substrates used for the TEGs. The process was run in $N_2$ atmosphere at a reduced pressure of < 1 hPa. The deposition was started with the growth of 10 nm $Al_2O_3$ layer at a low temperature, i.e. 50°C. The low temperature $Al_2O_3$ creates a barrier to prevent diethylzinc (DEZ) diffusion into the substrate [17]. Subsequently, the reactor was heated up to 200 °C for AZO deposition. The number of precursor cycles was calculated based on the target thickness and required amount of Al (2-4 %) in the final coating layer. Altogether 18 master cycles, i.e. 18 layers of ZnO followed by $Al_2O_3$ were deposited. Finally, the chamber was cooled down to 50 °C before unloading.

The second ALD process step with the same target thickness was carried out for the folded samples at 125°C due to the temperature limitation of the glue used in the folding procedure. From six to eight folded samples were placed in a mesh-basket for one AZO batch deposition. The mesh-basket enabled 3D coating and a conformal layer on each surface. Since the growth rate for ZnO is higher at 125°C than at 200°C, the deposition consisted of 16 master cycles instead of 18. The bottom $Al_2O_3$ layer was omitted to facilitate a good contact of the new AZO layer with the electrodes and AZO underneath. After deposition, the chamber was cooled down to about 50-60 °C before removing the samples from the reactor.

Silver electrodes were deposited on AZO by two different methods: screen printing from silver ink (Inkron IPC-114) and by electron beam evaporation from elemental silver (99.99 % purity). The latter method was used to analyse the specific contact resistance through transmission line model measurements, with eight 1 × 29 mm$^2$ contact pads deposited onto AZO in 6 mm intervals [18]. E-beam evaporated Ti/Au contacts were used for thermoelectric and Hall characterization of the films. The Seebeck coefficient of AZO was characterized with Linseis LSR-3 measurement system in helium atmosphere by gluing 4 × 20 mm$^2$ samples onto glass carrier chips with a ceramic adhesive (PELCO, Product No. 1606). Five thermal gradients were imposed over the samples and the Seebeck coefficient was extracted from a linear fit of the respective thermovoltages. The Hall characterization was carried out on 11 × 11 mm$^2$ samples in a van der Pauw four point contact geometry with the Ecopia HMS-5300 measurement system. Finally, the thermal conductivity of the films was measured using nanosecond transient thermoreflectance (Linseis) as reported previously [19]. In short, AZO grown on silicon is deposited with a 20/200 nm Ti/Au transducer layer. The transducer film is heated via an 8 ns laser pulse (Nd:YAG, 1064 nm), and its temperature dependent reflectance is monitored using a 463 nm diode pumped solid-state laser. The measured temperature transient is fitted to a one-dimensional transmission-line model using the bulk thermophysical parameters of the constituent films [20].

*2.2 Fabrication and characterization of the folded thin-film TEGs*

The process flow for fabricating the folded thermoelectric generator (TEG) modules is shown in Fig. 2. First, the yellow polyimide films of 25 μm in thickness are coated with ~400 nm thick AZO on which the silver ink electrodes are patterned by screen printing. The thermoelectric legs are then defined by mechanically removing AZO between the adjacent electrodes that will be at the same temperature (the positions indicated by black dashed lines in Fig. 1b). After defining, the TE elements on the substrate are connected electrically in series, while the folding of the substrate appropriately leads to a three dimensional structure where the TE elements are connected thermally in parallel as presented in Figs. 1 and 2 and described in more detail in Refs. [14, 15]. Another AZO coating is grown on the folded TE module by ALD to eliminate the influence of the cracks in the first AZO layer formed during the folding process. It should be noted, however, that the sharpest bends are imposed on the wide Ag conductor lines instead of the relatively brittle AZO. Finally, the second AZO layer is removed between the adjacent electrodes to maintain the correct electrical configuration. In the final configuration of the folded TEG, both the electrical current and heat flux are in the plane of the folded substrate, while simultaneously the temperature gradient ($\Delta T$), or heat flux, is perpendicular to the surface of the module (see Fig. 1b and Refs.[14, 15]).

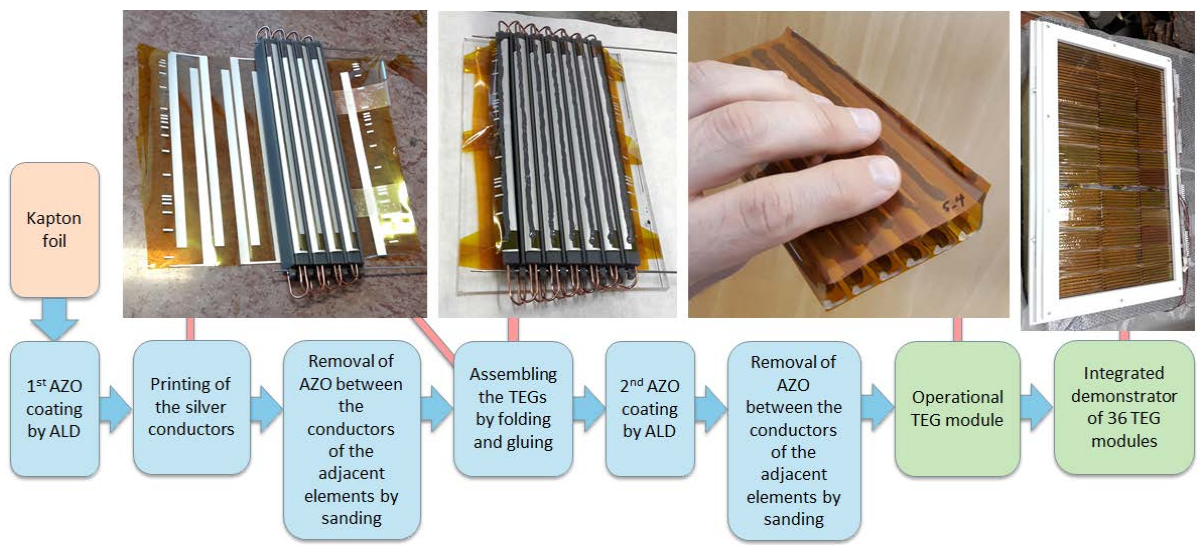

**Figure 2.** Process flow for fabricating the TEG modules.

The performance of the TEG module was characterized with a setup consisting of a temperature controlled heat plate under the TEG and ice on the top to cool the upper surface. An aluminium plate was applied between the module and the ice to distribute the heat evenly over the TEG area. Pt-100 temperature sensors integrated in the aluminium plate were used to measure the temperature of the cold side. To evaluate the performance, the TEG system was measured with different load resistances and the produced DC voltage was recorded as a function of the resistance. The voltage was measured with Fluke 111 multimeter with its own spike electrodes and the varied loading resistor was connected between its terminals. The DC power was calculated based on the measured values.

*2.3 Computational analysis*

For predicting the performance and extracting the material parameters of the folded TEG modules at the different fabrication stages, their operation was simulated using computational models based on the finite element method (FEM) realized in Comsol® Multiphysics. In the model, heat transfer equations are coupled with the electrical phenomena for modelling the thermoelectric effect (Peltier-Seebeck-Thomson) [21]. In addition to the FEM model, analytical methods were applied for calculating the selected characteristics of the TEGs. The model and the computational procedures are described in more detail in Refs. [14, 15]. In order to extract the contact resistance between AZO and silver ink, the sheet resistances of both the materials were measured and the corresponding electrical conductivities ($\sigma$) used as input parameters in the simulations. The contact resistance of the model was then adjusted to reproduce the measured resistance of the TEG module. Similarly, the Seebeck coefficients of the fabricated TEG modules were extracted by fitting the simulated results to the maximum measured output power for specified temperature gradients.

**3. Results**

*3.1 Material characterization and stability studies*

The electrical and thermoelectric properties of AZO were measured for the films of 380-400 nm in thickness grown on planar Kapton sheets at 200 °C. Depending on the sample, the measured conductivity varied roughly from 2 to $5 \times 10^4$ S/m (i.e. sheet resistance from 50 to 120 $\Omega$), while the Seebeck coefficient, *S*, and power factor. $PF = \sigma S^2$, were within $70 - 82$ $\mu$V/K and $0.8 - 1.2 \times 10^{-4}$ W/m/K$^2$, respectively.

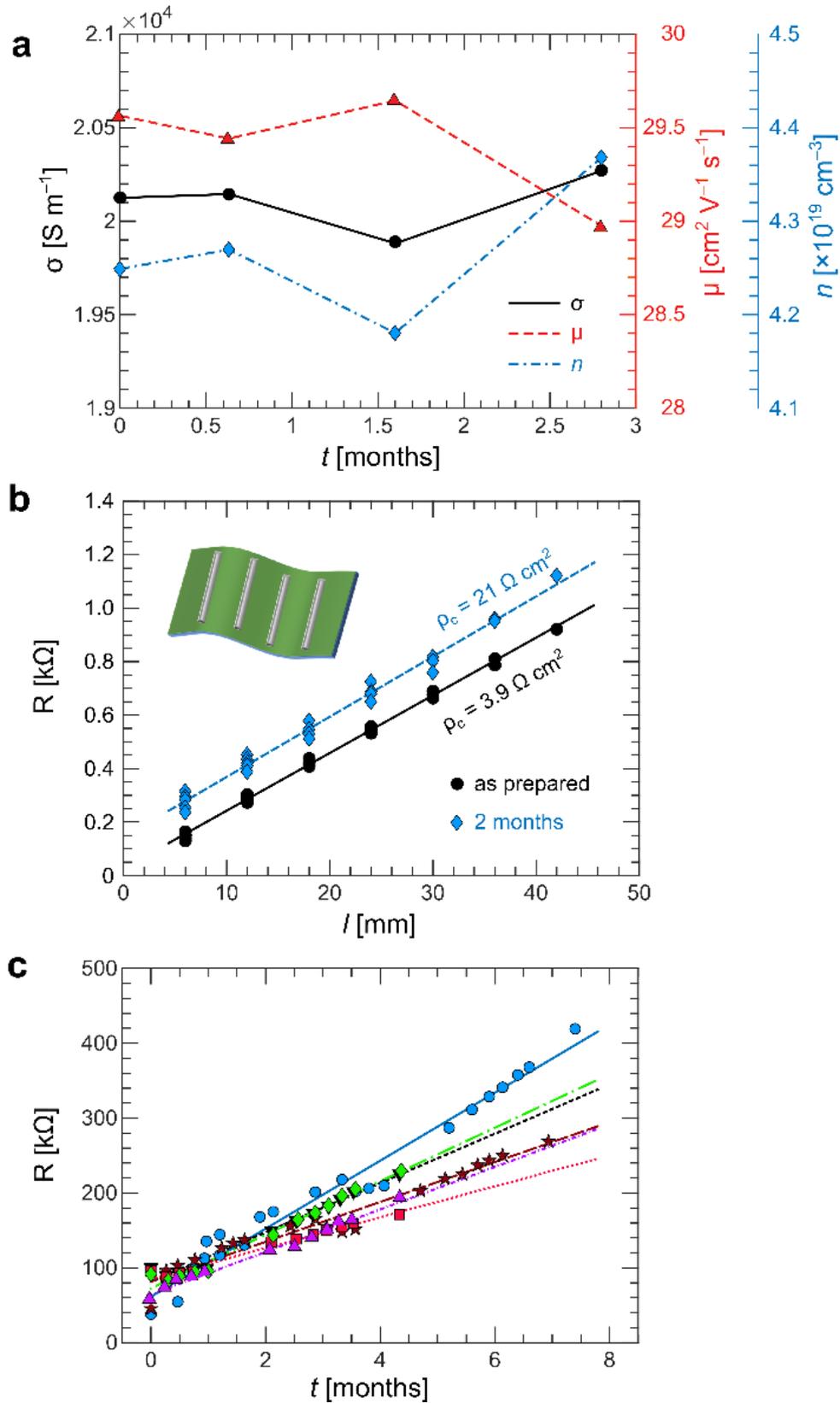

**Figure 3.** Stability studies of the electrical properties: (a) Hall characterization of AZO as a function of time, (b) transmission line measurement on evaporated silver as prepared and after two months (inset: contact geometry), and (c) the resistance of six TEG modules having ten TEG elements in series as a function of time.

To assess the stability of the electrical properties of the materials in ambient atmosphere, the mobility, $\mu$, conductivity, $\sigma$, and charge carrier concentration, $n$, of AZO, as well as the contact resistance, $\rho_c$ were monitored over a span of several months as shown in Figs. 3a and 3b, respectively. The Hall characteristics of the AZO were found highly stable within 5 % from the initial values, suggesting suitability for operation over extended periods. Consistently, the transmission line measurements carried out over the AZO film showed negligible change in the sheet resistance, as evidenced by the almost identical slopes of the lines shown in Fig. 3b. In contrast, the specific contact resistivity calculated from the extrapolated resistance at the zero contact separation ($l = 0$), showed over a five-fold increase from 3.9 up to 21 $\Omega$ cm$^2$ over a period of two months. Thus, while silver is the most broadly utilized material for printed conductors, the stability of the contact between silver and AZO requires addressing for long-term device operation. Such trend further translated into the fabricated large-area TEG modules: Fig. 3c shows the measured series resistance of the TEG modules as a function of time. The two-point resistance measurement with spike electrodes was carried out for six modules over ten elements of the twelve on each foil while keeping the modules at room temperature. The series resistance was between 40 and 100 $\Omega$ in the beginning, i.e. shortly after finishing the fabrication of the module as described in Fig. 2. During the measurement periods of several months, the series resistance increased at the rate of about 1 $\Omega$ per day. Similar trend of increased resistance was also observed in the planar modules with printed silver ink contact lines, while the measured sheet resistance of AZO and silver conductor remained the same over time. This suggests that the gradual resistance increase in the folded modules is not due to an increase in the resistivity of the AZO or silver ink materials generated by a tension in the folded shape but most likely relates to the contacts between silver and AZO. The increase results purely from the aging-related processes.

*3.2 Power production properties of the folded thin-film TEGs*

Fig. 4. shows the simulated temperature gradients ($\Delta T$) sustained by different materials under heat sink limited conditions as a function of the thickness of the material or TEG module. The calculations assume a constant temperature on the "hot" side (33 °C) and that the temperature of the air on the "cold" side of the module is 23 °C. Free convection is used for heat transfer between the air at 23 °C and the "cold" TEG surface. These assumptions result in the so-called heat sink limited conditions, setting a constraint to the maximum $\Delta T$ that can be maintained over the TEG module [16]. For comparison, the data is also shown for air, a polymeric material and a common building red brick, in addition to the thermoelectric device structures. The data indicates that the thermal transfer is very small in the folded module (open square) compared to the conventional TEGs (solid squares) where the heat flows across the plane of the TE films together with the electrical current. In fact, the thermal transport in the folded TEG is very similar to that in air. It can be seen that the conventional TEGs can only sustain a fraction of the temperature gradient

maintained by the folded TEG module in the heat sink limited mode where only the natural convection is taking care of the heat transfer from the cold surface of the module. In fact, the TE material of the folded module could be significantly thicker (open circles in Fig. 4) and still the folded TEG would maintain the advantage of low thermal conductivity and high $\Delta T$ as compared with the conventional TEGs.

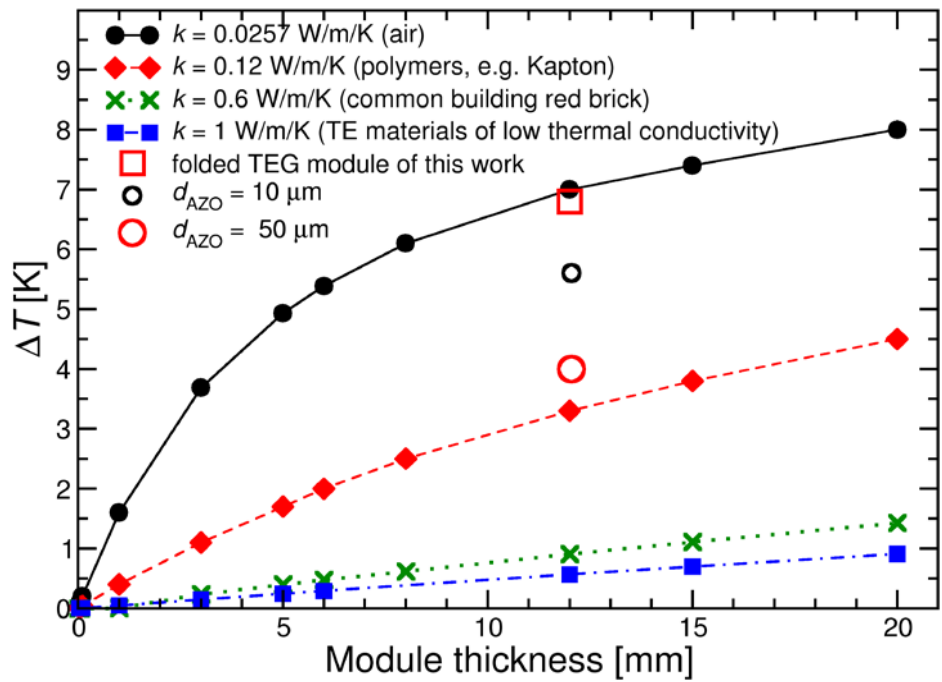

**Figure 4.** Temperature gradients ($\Delta T$) sustained by different materials or modules under heat sink limited conditions as a function of thickness (solid symbols) compared to that of the folded module of this work (open square; for parameters, see Table 1) and to the similar folded modules but with thicker AZO layers (open circles; $d_{AZO}$ is the thickness of AZO). The temperature gradient was initially 10 K: the temperature of the "hot" surface was set to 33 °C and the air on the opposite ("cold") side to 23 °C. Free convection was assumed between the air of 23 °C and the TEG surface on the "cold" side (the heat transfer coefficient was 5 W/m²/K). $k$ is the thermal conductivity.

The thermoelectric properties of a single folded TEG module covering an area of ~87 cm² (shown in Fig. 1c) were studied for two different temperature gradients. For $\Delta T$ = 49 K, the module consisting of eleven elements connected in series, produced the maximum power of 1.2 µW with the load resistance value equal to the internal series resistance of the module (~100 Ω). For $\Delta T$ = 10 K, i.e. for $\Delta T$ that is more often occurring in our everyday environment e.g. in buildings, the device was able to generate a power of 0.055 µW, corresponding to 3.2 µW for a 0.5 m² window.

The DC power generated by 36 TEG modules of eleven elements covering an area of 0.33 m² (as shown in the rightmost photograph of Fig. 2) is depicted in Fig. 5 as a function of load resistance for both measured and simulated data. The results for a TEG measured in its planar configuration is also included by assuming it to be folded and multiplied to cover the same area as the 36 folded modules. The contact resistance between the silver ink conductor lines and AZO was ~30 $\Omega cm^2$, as extracted from the FEM simulations fitted to the experimental results with the resistance value of 100 $\Omega$ measured for a single TEG module. As can be seen, a good fit between the experimental and simulated data was obtained with the mean relative error being 3.3 %, 10 % and 13 % for the dashed, dash-dotted and solid blue curves, respectively. For comparison, additional simulated data is shown for a case where AZO is replaced by $Bi_2Te_3$ in a similarly folded TEG assembly of 36 modules. The data includes two different contact resistance values between $Bi_2Te_3$ and the silver conductor lines, assuming a good contact (0.01 $\Omega cm^2$) and a contact resistance similar to the one between AZO and silver ink (30 $\Omega cm^2$). The material parameters of $Bi_2Te_3$ were taken from Ref. [22], while the other parameter values used in the simulations of Fig. 5 are listed in Table 1. Furthermore, simulated data is shown for the folded modules with 50 $\mu$m thick AZO (with $S$ = 73 $\mu$V/K) and $Bi_2Te_3$ layers in order to predict the potential for improvement in the power production with a thicker TE material caused by the decreased resistance.

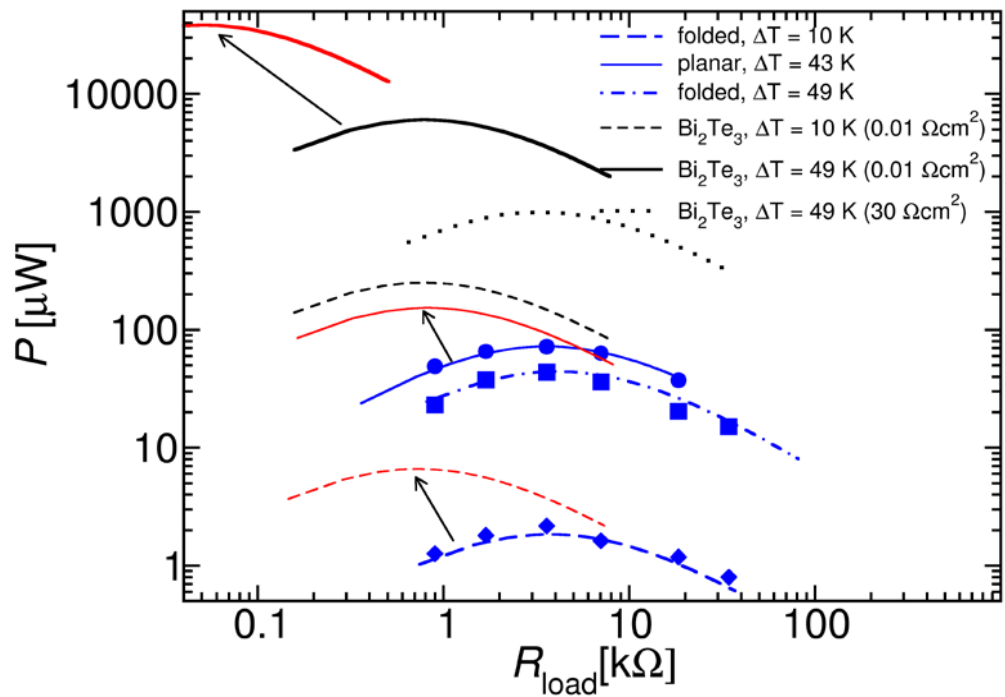

**Figure 5.** The DC power generated by 36 TEG modules covering an area of 0.33 $m^2$ as a function of load resistance. Simulated data is also shown for $Bi_2Te_3$ with two different contact resistances to the conductor lines, assumed to replace AZO in the folded TEGs. Measured (markers) and simulated (lines) results. The arrows indicate the shift of the corresponding curves if the TE materials were replaced with the corresponding TE material of 50 $\mu$m in thickness (red curves).

**Table 1.** Parameters used in the device simulations.

| Parameter | AZO | Silver | Kapton |
|---|---|---|---|
| Sheet resistance [$\Omega$ sq$^{-1}$] | 53 | $7.5 \times 10^{-4}$ | |
| Thickness for planar/folded [µm] | 0.38/0.76 | 50 | 25 |
| Thermal conductivity [W m$^{-1}$ K$^{-1}$] | 3.6 | 238* | 0.12* |
| Electrical conductivity for planar/folded, $\sigma$ [S m$^{-1}$] | $5\times10^4$ / $3\times10^4$ | $2.67 \times 10^7$ | |
| Contact resistance [$\Omega$cm$^2$] | 30** | | |
| **Before folding:** | | | |
| Seebeck coefficient, $S$ [µV K$^{-1}$] | −73** | 3.5* | |
| Power factor, $\sigma S^2$ [W m$^{-1}$ K$^{-2}$] | $2.7 \times 10^{-4}$ | | |
| **After folding:** | | | |
| Seebeck coefficient, S (µV K$^{-1}$) | −48** | 3.5* | |
| Power factor, PF (W m$^{-1}$ K$^{-2}$) | $1.2 \times 10^{-4}$ | | |

Origin of the data: Measured or calculated from the measured/fitted values.
*From literature [23, 24]. **Obtained by fitting the simulations to the experimental data.

## 4. Discussion

It was shown that the planar TEGs produce somewhat higher power even for slightly lower $\Delta$T = 43 K when compared to the folded modules for $\Delta$T = 49 K (Fig. 5). The result is in accordance with the extracted material parameters of AZO shown in Table 1: AZO in the folded TEGs exhibits lower values for both the Seebeck coefficient and electrical conductivity. This may relate to the lower growth temperature used in the second ALD process of AZO on the folded TEG or to the fact that the second ALD process may not be able to fully recover the influence of the cracks formed in the lower AZO layer during the folding process, or to both.

The data calculated for $Bi_2Te_3$ elucidates the power levels achievable with the folding density (corresponding to the number of TEG elements per unit area) implemented in the present study of the large-scale system, if a high-performance room temperature thermoelectric material were used. In principle, it should be possible to increase the folding density without increasing significantly the heat leakage through the module, as shown in Ref. [14], implying that the power output can be still increased above the solid black line of Fig. 5, if high-performance thermoelectric materials are available. Although not feasible with ALD, the positive influence of simply increasing the thickness of the TE layers of AZO and $Bi_2Te_3$ without changing the folding density is also demonstrated in Fig. 5.

To estimate the usefulness of the produced power in relation to practical use scenarios, the power requirements of a wireless sensor node designed for monitoring environmental variables in a building are considered. The electronics module of the sensor node consists of an energy harvesting circuit with a storage capacitor between the TEGs and the power consuming electronics, an integrated microcontroller and a Bluetooth LE radio transceiver unit, as well as temperature, humidity and $CO_2$ sensors, as described in Ref. [15]. Assuming temperature and humidity measurements and data transmission every 15 minutes, the average power consumption of the electronics is 3.3 µW. If a $CO_2$ measurement once an hour is added to the operation, the average power consumption is increased to 32.5 µW. The area of the folded TEGs of the present study required to produce the power of the first use scenario is 0.024 m$^2$ for $\Delta T =$ 49 K while the corresponding area needed to satisfy the power requirements of the second use scenario of 32.5 µW is 0.23 m$^2$. If the folded TEGs were made of $Bi_2Te_3$, the required areas would reduce to 0.0002 m$^2$ and 0.0019 m$^2$ for the two use scenarios for $\Delta T =$ 49 K and 0.004 m$^2$ and 0.042 m$^2$ for $\Delta T =$ 10 K, respectively.

The most significant cost reduction compared to the conventional TEGs is expected to come from the materials costs and from the fact that no additional heatsinks are needed. AZO is significantly less expensive than the typical conventional TE materials (2.3 US$/kg for AZO versus e.g. 110 US$/kg for $Bi_2Te_3$ [25]) and the consumption per unit area is small for the thin films. The amount of AZO in the TEG module of this work is ~120 mg. Polymeric substrates can also provide cost-effective solutions for the folded TEGs, although the Kapton foil used in this study is relatively expensive and its costs are clearly more than the chemical costs for coating. In the case of using ALD, the overall expenses come from heating the ALD reactor (chamber, pipes), use of the pump and it's cooling (water), electronics overall, purging gas (N2), chemicals (DEZ, TMA), etc. The tool time and personal effort are also much more costly than the AZO coating layer itself. The realization of a truly low cost folded TEG requires a carefully considered production process together with the right substrate material.

The results suggest that useful power can be produced with the developed TEG concept using AZO. The increase in the contact resistance between AZO and silver ink as a function of time is, however, still an issue that needs special attention before the proposed system fulfils the demands of commercial applications. The literature reveals that silver has been used to form both Ohmic [26] and Schottky contacts to ZnO [27]. In this connection, the sensitivity of the contact to different surface conditions of ZnO and the tendency of silver to oxidize easily have been reported. The observed instabilities in the contact resistances may be of similar origin. However, further studies are still required. An option is to look for more suitable material combinations and fabrication methods to eliminate the issues with the contact resistance as well as the limitations set by the tendency of AZO for cracking on flexible substrates.

## 5. Conclusions

A new folded thin-film thermoelectric generator has been implemented and the concept demonstrated for large area applications. The demonstrated TEG is based on a transparent AZO thin-film deposited on a flexible substrate. It was shown that functional TEGs for large areas can be fabricated based on the folded TEG concept. The proposed concept has clear advantages over the available commercial TEGs under heat sink limited conditions, although the materials used for the demonstration still have some issues concerning the long-term stability and mechanical durability. The results indicate that the thermal leakage in the folded TEG structure is minimal compared to the commercially available bulk or thin-film TEGs in which the heat flux and electrical current are perpendicular to the plane of the TE films. In low-energy-density applications, especially when heat sink limited operation is required, i.e. when only free convection can be used for transferring heat away from the cold surface of the TEG, the folded TEG exhibits superior performance by enabling significantly higher temperature gradients for power production than the conventional TEGs. Thus, although the developed folded TEGs require relatively large areas for power production, they are lightweight (and translucent in case of AZO) and can operate also under such heat sink limited conditions where the conventional TEGs are practically useless. In fact, the demonstrated TEG concept is believed to be most beneficial in environments where significant temperature gradients are available, light weight is an advantage and efficient heat sinks cannot be used, such as airplanes.

Declarations of interest: none.

**Acknowledgements**

This work was supported in part by the European Union's Horizon 2020 Research and Innovation Program 2014–2018 as part of the TransFlexTeg project under Grant 645241 and in part by the VTT Technical Research Centre of Finland Ltd. The authors would also like to thank their colleagues K. Kiri, R. Grenman, and M. Vilkman for screen printing the silver conductors on the TEG sheets.